\documentclass[3p,lefttitle]{elsarticle}
\usepackage{graphics}
\usepackage[T1]{fontenc}
\usepackage{amsmath,amsfonts,amsthm}
\usepackage[utf8]{inputenc}
\usepackage{newpxtext,newpxmath}
\usepackage[normalem]{ulem}
\usepackage{color}
\usepackage[normalem]{ulem}
\usepackage{amsmath}
\newcommand{\stkout}[1]{\ifmmode\text{\sout{\ensuremath{#1}}}\else\sout{#1}\fi}

\begin{document}
\def\parsedate #1:20#2#3#4#5#6#7#8\empty{20#2#3/#4#5/#6#7}
\def\moddate#1{\expandafter\parsedate\pdffilemoddate{#1}\empty}
\bibliographystyle{apsrev}
\title{Monitoring Decoherence via Measurement of Quantum Coherence}
\author[1]{Anu Venugopalan}
\ead{anu.venugopalan@gmail.com}

\author[1]{Sandeep Mishra}
\ead{sandeep.mtec@gmail.com}

\author[2]{Tabish Qureshi}
\ead{tabish@ctp-jamia.res.in}

\address[1]{University School of Basic and Applied Sciences, G.G.S. Indraprastha University, Sector 16C Dwarka, Delhi - 110078, India.}
\address[2]{Centre for Theoretical Physics, Jamia Millia Islamia, New Delhi-110025, India.}

\begin{abstract}
A multi-slit interference experiment, with which-way detectors, in the
presence of environment induced decoherence, is theoretically analyzed.
The effect of environment is modeled via a coupling to a bath of harmonic
oscillators. Through an exact analysis, an expression for $\mathcal{C}$, a
recently introduced measure of coherence, of
the particle at the detecting screen is obtained as a function of the
parameters of the environment. It is argued that the effect of
decoherence can be quantified using the measured coherence value which lies
between zero and one. For the specific case of two slits, it is shown that
the {\em decoherence time} can be obtained from the measured value of the
coherence,  $\mathcal{C}$   {, thus providing a novel way to quantify the
effect of decoherence via direct measurement of quantum coherence. This
would be of significant value in many current studies  that seek to
exploit quantum superpositions for quantum information applications and
scalable quantum computation.}
\end{abstract}


\maketitle

\section{Introduction}

 The dynamics of a quantum system weakly coupled to a large number of 
degrees of freedom, the 'environment', has been a much studied subject. Its proposed connection with the emergence of classicality, led to the very active field of decoherence \cite{zeh,zurek,max}. The central idea of the decoherence approach has been that 'classicality' is an emergent property of systems interacting with an environment which `washes away' quantum coherence. The qualitative and quantitative study of decoherence has provided valuable insights into the actual mechanism of
the loss of quantum coherences and some of its predictions  have also been
successfully tested  experimentally \cite{exptdecoh1,brune, monroe}. As decoherence
would naturally ruin delicate quantum features and hence the functioning
of devices which use quantum coherence for information processing,
its study is  highly relevant to all experimental implementations of
quantum information and computation \cite{monroe,qcompdec1,qcompdec2,qcompdec3}.

The essential idea of decoherence is the following. Since entanglement is
a generic outcome of most interactions, a quantum system interacting with
the environment gets entangled with certain environment states. In such
a situation the quantum system can then be sensibly described only by a
reduced density operator, by tracing over the states of the environment \cite{zurek82}. In this sense, an initial pure state constructed as a coherent
superposition decoheres into a statistical mixture when its dynamics
incorporates the coupling to a large number of environmental degrees of
freedom. While the pure state density matrix of the system (when viewed
in a particular basis) has both diagonal and off-diagonal elements,
it can be seen that after a certain time, impacted by environmental
influence, the off-diagonal elements (which reflect quantum coherence)
of the reduced system are diminished to give a statistical mixture.
In the extreme case the reduced density matrix of the system may become completely diagonal in a particular basis. In such a situation the system is said to have fully decohered, and its coherence would be completely lost. The degree of decoherence  undergone by a system clearly seems to be intimately connected to the coherence remaining in the system. Taking a cue from this, in the following we will
explore the decoherence of a system by looking at its remaining coherence.

Recently a new measure of coherence was introduced in the context of
quantum information theory, which is just the sum of the absolute values of the off-diagonal elements of the density matrix of a system, namely
$\sum_{i\neq j} \langle i | \rho | j \rangle$, where $|i\rangle, |j\rangle$
are states of a particular basis set \cite{coherence}.
As is obvious, this measure is basis dependent, and has the minimum
value zero, for a diagonal density matrix. However, there is no 
well-defined upper limit to this measure, as it depends on the
dimensionality of the Hilbert space of the system. A measure with a 
well-defined upper limit is desirable.
Using the measure of Baumgratz, Cramer, and Plenio, a normalized (basis dependent) quantity called coherence was also recently introduced \cite{nduality}
\begin{equation}
{\mathcal C} = {1\over n-1}\sum_{i\neq j} |\langle i|\rho|j\rangle|,
\label{coherence}
\end{equation}
where $n$ is the dimensionality of the Hilbert space, and the value of
$\mathcal{C}$  always lies between 0 and 1. It is straightforward to
see that a completely decohered system, corresponding to a fully diagonal
density matrix, has coherence $\mathcal{C} = 0$. One can verify that 
$\mathcal{C}$ will take value 1 for a pure density operator corresponding to
the following maximally coherent state
\begin{equation}
|\psi\rangle = \frac{1}{\sqrt{n}} \sum_{k=1}^n |k\rangle.
\end{equation}

In the present investigation we theoretically analyze a multi-slit
interference experiment by including the effect of
environment on the particle passing through the multi-slit.
For multi-slit interference it was recently demonstrated that
coherence can be experimentally measured \cite{tania}.
Apart from this, coherence in a multi-slit experiment can be related to
measurable quantities in various ways \cite{biswas}. In this paper we
show that incorporating the interaction with the environment modifys
the  position space probability distribution (interference pattern)
and the expression for quantum coherence in an interesting way, clearly
illustrating the decoherence mechanism.
For the simplest case of two
slit interference, this allows us to calculate the decoherence time
using quantum coherence measurements.

\begin{figure}
\centerline{\resizebox{12cm}{!}{\includegraphics{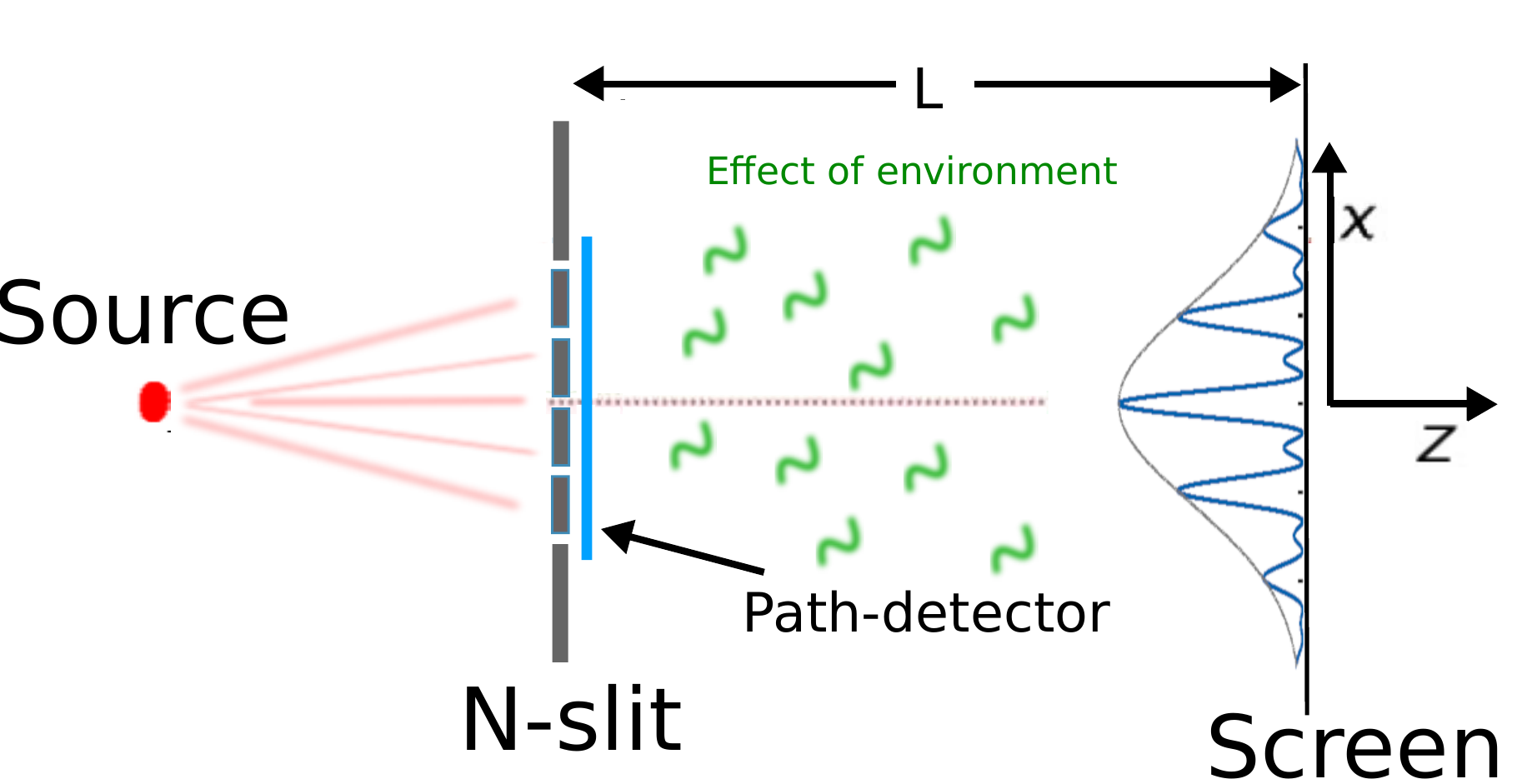}}}
\caption{Schematic diagram of a n-slit interference experiment, with 
a quantum path-detector. The interfering particle is affected by  
interaction with an environment.
}
\label{nslit}
\end{figure}

\section{Multi-slit interference}
\subsection{Interference and path-detection}

To set the ball rolling, let us first consider a quantum particle
(quanton) passing through
$n$-slit{s}. If $|\psi_i\rangle$ represents the amplitude
of the particle to pass through the $i$'th slit, then the state of
the particle, after passing through the {$n$} slit{s}, can be described as a 
superposition of all possible amplitudes, i.e., $|\psi_i\rangle$, to go through different slits.  The states $|\psi_i\rangle$ are mutually orthogonal by virtue of the slits being narrow and spatially separated. We choose $|\psi_i\rangle$ to be
normalized, and associate a weight factor with it, which determines the
probability of the particle to go through a particular slit:
\begin{equation}
|\Psi_0\rangle = c_1|\psi_1\rangle + c_2|\psi_2\rangle +
\dots , c_n|\psi_n\rangle .
\label{pure}
\end{equation}
 {$|\Psi_0\rangle $ is normalized, and the sum of the probabilities
 $\sum_{i=1}^n|c_i|^2=1$.}
The probability density of the particle hitting the screen at a particular
position {$x$ on the screen (Fig.1) }is given by $|\langle x|\Psi_0\rangle|^2$. The pattern on the screen has the following general form
\begin{equation}
|\langle x|\Psi_0\rangle|^2 = \sum_{i=1}^n|c_i|^2|\langle x|\psi_i\rangle|^2 
+ \sum_{j\neq k} c_j^*c_k\langle x|\psi_k\rangle\langle \psi_j|x\rangle.
\label{patternp}
\end{equation}
The first term represents the sum of patterns formed by the particles
coming out of individual slits, as if the other slits did not exist.
The second term represents the interference between the amplitudes of particle
coming out of j'th and k'th slits, summed over all j's and k's which are 
different.  Clearly, the multi-slit interference pattern is built up of
all possible two-slit inteference terms ($j\neq k$). 

Since we are interested in probing which slit the particle went through,
we need to have some kind of detector which does this job. We assume that
our detector is the simplest, and is fully quantum mechanical. In the von Neumann scheme,  measurements are described by treating both
the system and the measuring apparatus (detector) as quantum objects and
for  a measurement to be affected, the measured system interacts with
the detector  to produce an entangled state with one-to-one correlations
between the system and the detector states \cite{neumann}.

Thus, in order for the path-detector to be capable of detecting which of the
$n$ paths the particle took, an entangled state
of the following kind should result
\begin{equation}
|\Psi\rangle = c_1|\psi_1\rangle |d_1\rangle + c_2|\psi_2\rangle  |d_2\rangle +
\dots , c_n|\psi_n\rangle |d_n\rangle,
\label{ent}
\end{equation}
where $|d_i\rangle$ is the state of the path-detector if the quanton
went through the $i$'th path.
We choose the detector states
$\{|d_i\rangle\}$ to be normalized, but not necessarily orthogonal.
With the path-detector added to the interference setup, the pattern
of the particles hitting the screen gets modified, and has the following form
\begin{equation}
|\langle x|\Psi_0\rangle|^2 = \sum_{i=1}^n|c_i|^2|\langle x|\psi_i\rangle|^2 
+ \sum_{j\neq k} c_j^*c_k\langle x|\psi_k\rangle\langle \psi_j|x\rangle
\langle d_j|d_k\rangle .
\label{patternr}
\end{equation}
 From (\ref{patternr}) one can see that the first term remains unaffected by the introduction of path-detector.
This is obvious as the presence of a path detector is not expected
to affect the probability of the particle to pass through a single slit.
The second term, which gives rise to interference, however, is reduced by the
factors $\langle d_j|d_k\rangle$. If the path-detector states are completely
orthogonal, $\langle d_j|d_k\rangle = 0$, it is clear from
(\ref{patternr}) that  the interference
term disappears. If the path-detector states are all identical,
i.e., $\langle d_j|d_k\rangle = 1$ for all $j,k$, (\ref{patternr}) reduces
to (\ref{patternp}).
This is in tune with the understanding that any attempt at gaining
information about which slit the particle went through degrades the interference,  while a complete ignorance about which-path information preserves the interference.

\subsection{Effect of the environment}

Let us now assume that as the particle (quanton) comes out of the
n-slit{s} and travels
to the screen, it is affected by weak interaction with some kind
of environment. We describe this environment as a reservoir of
non-interacting quantum oscillators, each of which interacts with the
particle. The Hamiltonian
governing the particle can then be represented as
\begin{equation}
H = \frac{p^2}{2m} + \sum_j {P_j^2\over 2M_j} 
    + {1\over 2}M_j \omega_j^2\left(X_j-{g_jx\over {M_j}\omega_j^2}\right)^2,
\end{equation}
where $x,p$ are the position and momentum operators of the particle {(quanton)},
$m$ is its mass, 
$X_j,P_j$ are position and momentum operators , and $M_j$ the mass
of the {jth} harmonic oscillator of frequency $\omega_j$
comprising the environment and $g_j$ are the respective coupling strengths.
The dynamics of a particle in simple potentials and in interaction with the
environment modelled as a harmonic oscillator heat bath has been studied at great length in recent literature
\cite{legget,gsa,dekker,barchielli,dk, av}.  The dynamics for the closed
combine of the quanton and the environment is governed by the Hamiltonian
evolution via (7) and the Schr\"{o}dinger equation. A tracing over
all the degrees of freedom of the environment results in an equation
describing the dynamics of  the {\em reduced density matrix} of just
the quanton. This reduced density matrix evolves according to a master
equation which is obtained by solving the Schr\"{o}dinger equation for the
particle and the environment and then tracing over the environment degrees
of freedom. Several authors have worked extensively on the decoherence
approach using the master equation for the reduced density matrix. The
master equation for this kind of model of the environment was first
derived separately by Caldeira and Leggett \cite{legget}, Agarwal \cite{gsa}, Dekker \cite{dekker} and others in the context of quantum Brownian motion and
is popular in  the study of open quantum systems. For our purpose, we
deal with the  master equation for the reduced density matrix for the
particle (quanton) for the particle-environment composite described by
(7), after the environment degrees of freedom are traced out:
\begin{eqnarray}
\frac{\partial\rho(x,x',t)}{\partial t} &=& \left\{\frac{-\hbar}{2im}
\left( \frac{\partial^2}{\partial x^2}
- \frac{\partial^2}{\partial x'^2} \right) \right.\nonumber\\
&&\left. -\gamma(x-x')\left( \frac{\partial}{\partial x} - \frac{\partial}{\partial x'} \right) 
- \frac{D}{4\hbar^2}(x-x')^2\right\}. \nonumber\\
\label{master}
\end{eqnarray}

Here $\rho(x,x',t)$ is the reduced density matrix of the particle
 in the position degrees of freedom, $\gamma$ is the Langevin friction
 coefficient and $D = 2m\gamma k_BT$ can be interpreted as
 a diffusion
coeficient, and $T$ is the temperature of the harmonic oscillator
heat-bath {\citep{legget,dekker, gsa} . $\gamma$ and $D$ are related to the parameters of the total Hamiltonin (7). This master equation can be seen to naturally separate into three terms, one representing pure quantum evolution, one leading to dissipation or relaxation,  and one causing diffusion. It has been
widely reported in the literature that in the dynamics governed by
such a master equation, coherent quantum superpositions persist for
a very short time as they  are rapidly destroyed by the action of the
environment. It is generally  agreed that the two main features seen as
signatures of decoherence are : (a) the decoherence time, over which the
superpositions decay is much much shorter than any characteristic time
scale of the system (the thermal relaxation time, $\gamma^{-1})$, and
(b) the decoherence time varies inversely as the square of a quantity
that indicates the `size' of the quantum superposition. This feature
has also been reported in experiments \cite{brune, monroe}.

In the following we will attempt to quantify the effect of environment
induce decoherence on a particle undergoing a n-slit inteference with
the possibility of which path detection.

\subsection{Decoherent dynamics of the particle}

For calculational simplicity we assume that the state $|\psi_k\rangle$ 
which emerges from the k'th slit, is a Gaussian wave-packet localized 
at the location of the k'th slit, with a width equal to the width of the
slit:
\begin{eqnarray}
\langle x|\psi_k\rangle &=& \tfrac{1}{(\pi/2)^{1/4}\sqrt{\epsilon}} e^{-(x-k\ell)^2/\epsilon^2},
\label{psik}
\end{eqnarray}
where $\ell$ is the distance between the centres of two neighboring slits,
and $\epsilon$ is their approximate width. 
Following an interaction of the particle (quanton) with the which-path
detector, the combined wave-function of the particle and the detector,
as it emerges from the n-slits,
is given by
\begin{eqnarray}
\langle x|\Psi\rangle &=& \tfrac{1}{(\pi/2)^{1/4}\sqrt{\epsilon}} 
\sum_{k=1}^n c_k e^{-(x-k\ell)^2/\epsilon^2}|d_k\rangle ,
\label{Psi}
\end{eqnarray}
which is a specific form of {the entangled state} (\ref{ent}).
The density matrix corresponding to this can be written as
\begin{eqnarray}
\rho_0(x,x') &=& \tfrac{1}{\sqrt{\pi/2}\epsilon}
\sum_{j,k} c_jc_k^* e^{\frac{-(x-j\ell)^2}{\epsilon^2}}
e^{\frac{-(x'-k\ell)^2}{\epsilon^2}}|d_j\rangle\langle d_k|.~~~~
\label{rho}
\end{eqnarray}
If one were to look only at the particle, it amounts to  tracing over
the states of the path detector, giving us the reduced density matrix
for the quanton
\begin{eqnarray}
\rho(x,x',0) &=& \tfrac{1}{\sqrt{\pi/2}\epsilon}
\sum_{j,k} c_jc_k^* e^{\frac{-(x-j\ell)^2}{\epsilon^2}}
e^{\frac{-(x'-k\ell)^2}{\epsilon^2}}\langle d_k|d_j\rangle.~~~~
\label{rho0}
\end{eqnarray}
This is the density operator of the particle at time $t=0$ as it emerges
from the n-slit. Its decoherent dynamics will be governed by equation
(\ref{master}). We assume that after a time $t$ the particle reaches the 
screen, after traveling a distance $L$, and its density operator is given
by $\rho(x,x',t)$, { described in the next section}.

\section{Results}
\subsection{Interference with decoherence}

The measured intensity on the screen after the  n-slits would be just the probability of
the particle hitting the screen at a particular position. This, in turn, 
corresponds to the diagonal terms of the reduced density matrix of the particle, in the position basis. Eqn. (\ref{master}) can be solved exactly, with the
initial
condition (\ref{rho0}), to yield $\rho(x,x',t)$. It's diagonal component,
$\rho(x,x,t)$,
representing the probability density of the particle hitting the screen
at a point $x$, is given by
\begin{eqnarray}
\rho(x,x,t) &=& \frac{1}{\sqrt{\pi \alpha/2}}\left[\sum_{j=1}^n |c_j|^2
e^{-{2(x-jl)^2\over \alpha}}\right. \nonumber\\
&&\left. + \sum_{j\ne k}
|c_jc_k||\langle d_k|d_j\rangle| e^{-\frac{1}{\alpha} f _{jk}(x)}\right. \nonumber\\
&&\left.\cos\left\{\tfrac{2\hbar(1-e^{-2\gamma t})\ell(k-j)
(x-\ell\frac{k+j}{2})}{\alpha\gamma m\epsilon^2} + \theta_k-\theta_j\right\}\right]\nonumber\\
\label{rho}
\end{eqnarray}
where\\
$f_{jk}(x)= (x-j\ell)^2 + (x-k\ell)^2 +
\frac{l^2(j-k)^2D
[4\gamma t+4e^{-2\gamma t}-e^{-4\gamma t}-3]}{16\epsilon^2m^2\gamma^3} $,\\
and
$\alpha = \epsilon^2 + \frac{\hbar^2(1-e^{-2\gamma t})^2}{\epsilon^2m^2\gamma^2}
+ \frac{D
[4\gamma t+4e^{-2\gamma t}-e^{-4\gamma t}-3]}{8m^2\gamma^3}$. {Also,
for convenience, we have combined the phases of $c_j$ and $|d_j\rangle$
as $c_j |d_j\rangle=
|c_j ||d_j\rangle e^{i \theta_{j}}$ , where $|d_j\rangle$ is now real.}
Eqn. (\ref{rho}) represents the exact dynamics of a particle passing 
through a multi-slit {with which path detection} and interacting with an environment. It can be used to
describe the dissipative dynamics of the particle. However, if one is
interested in studying the effects of decoherence, one is in the limit
of very weak coupling of the particle with the environment, where 
dissipative effects are negligible. Dissipation typically occurs at a
time-scale of $1/\gamma$. We assume that the effect of the environment is
so weak that dissipative time-scales are much longer than the time the
particle takes to reach the screen, $t \ll 1/\gamma$, or $\gamma t \ll 1$. {Further, it is assumed that the time evolution of the Gaussian can be calculated by either assuming the quanton to be a particle of mass $m$, moving with a momentum corresponding to a de Broglie wavelength $\lambda$,  or by assuming it to be a photon of wavelength $\lambda$ \cite{dillon}} .
We also assume the slit width to be very small, i.e., 
$\epsilon^2 \ll \lambda L/\pi$.
In this limit (\ref{rho}) reduces to
\begin{eqnarray}
\rho(x,x,t) &=& \frac{1}{\sqrt{\pi \alpha/2}}\left[\sum_{j=1}^n |c_j|^2
e^{-{2\epsilon^2(x-jl)^2/(\lambda L/\pi)}} \right.\nonumber\\
&& +\left. \sum_{j\ne k}
|c_jc_k||\langle d_k|d_j\rangle| e^{-\epsilon^2\{(x-j\ell)^2 + (x-k\ell)^2\}\over(\lambda L/\pi)^2}{e^{\frac{-D(j-k)^2 \ell^2 t}{12 \hbar^2}}}\right. \nonumber\\
&&\left.\cos\left\{\frac{2\pi\ell(k-j)
(x-\ell\frac{k+j}{2})}{\lambda L } + \theta_k-\theta_j\right\}\right],
\label{rhoa}
\end{eqnarray}
As a consistency check, we first consider the limit of coupling with the
environment going to zero, which  {is the regime when decoherence is
completely absent}. This is achieved
by taking the limits $\gamma\to 0,~D\to 0$. In this limit,
the term $e^{-D(j-k)^2 \ell^2 t/12 \hbar^2}$  in (\ref{rhoa})
becomes equal to unity,
and (\ref{rhoa}) reduces to eqn. (16) of ref. \cite{tania}. Thus in the limit
of  environmental {coupling}  going to zero, {we recover }known results for n-slit
interference. 

Let us now analyze the implications of the result (\ref{rhoa}) in somewhat 
greater detail. Note that the decohering n-slit interference pattern (\ref{rhoa})  is also built up from all possible two-slit inteference terms as was the case in (\ref{patternp}) when there was no coupling to the environment. The environmental coupling has had the effect of modifying these pair-wise contributions. The effect of decoherence is neatly condensed into
the exponential factor {$e^{-D(j-k)^2 \ell^2 t/12 \hbar^2}$}  multiplying
the cosine term
which gives rise to interference. It is evident that as time progresses,
the decoherence effect will degrade the interference. One would naively
think that the effect of environment will be to progressively
decrease the contrast of the interference, while retaining its overall
characteristic n-slit signature .
However, the interesting point to notice here is that the exponential
decay {term} in (\ref{rhoa})  cannot be pulled out of the summation,
{as} it depends on j, k. It is worth reminding that the {second summation term}  {over j, k, ($ j \neq k$)} in (\ref{rhoa})
represents the interference between the amplitudes from the j'th and
k'th slits. Here the argument of the exponent in the exponential
decay term is {$-D(j-k)^2 \ell^2 t/12 \hbar^2$}. Note that the
distance between two pairs of slits is $(j-k)\ell$. Let us now
understand the physical significance of this term. Clearly, the
larger the difference between j and k, the smaller is the magnitude
of the exponential decay term and hence the faster is the decay. This
means  that the decoherence effect has its strongest contributions
coming from the interference from pairs of slits which are the
farthest apart from each other and its weakest contributions
for interference from  neighboring pairs of slits which are next
to each other. This is quite clearly in consonance with the accepted
signatures of decoherence wherein  the decoherence time varies inversely
as the square of a quantity that indicates the `size' of the quantum
superposition \cite{brune, monroe}. In our case this 'size' is the distance
between the two slits, $(j-k)\ell$. Notice that for each pair of slits,
there will be a characteristic decoherence time, $\tau_{d}^{(jk)} =
12\hbar^2/D(j-k)^2 \ell^2$ and for n-slits there will be a
total of n(n-1)/2 time scales which will collectively contribute in the
summation leading to the degradation of the n-slit pattern. Of these
terms, the strongest contribution to the fast exponential decay will
come from one term corresponding to the pair of slits farthest apart. The
weakest contributions will come from (n-1) terms for pairs of neighbouring
slits and the coherence from their interfering amplitudes would survive
the longest.

For the simplest case of two slits, there will be only one
time scale, $\tau_{d} = 12\hbar^2/D \ell^2$}  =  $\frac{6\hbar^2}{m\gamma k_B T \ell^2}$.
A physical interpretation of this form of the decoherence time (leading
to  superpositions  vanishing on a time scale much shorter than the
relaxation time, $\gamma$) may be argued as  stemming from the fact
that in the model considered, the coordinate-coordinate coupling of the
particle to the environment is very weak and the states of the environment
correlated to the $n$ states of the particle emerging from the multi-slit,
$|\psi_1\rangle, |\psi_2\rangle, ...|\psi_n\rangle$ in
(3), are mutually orthogonal to each other \cite{legget}. Eq. (\ref{rhoa})
summarizes the central result of this work, describing the interference
pattern (position probability distribution)  for a quanton interacting
with an environment after emerging from n-slits with which way
detectors. This result can be applied to  recent experimental  studies
investigating  entanglement, quantum coherence and decoherence in matter
waves \cite{c60, neon}.

\begin{figure}
\centerline{\resizebox{12cm}{!}{\includegraphics{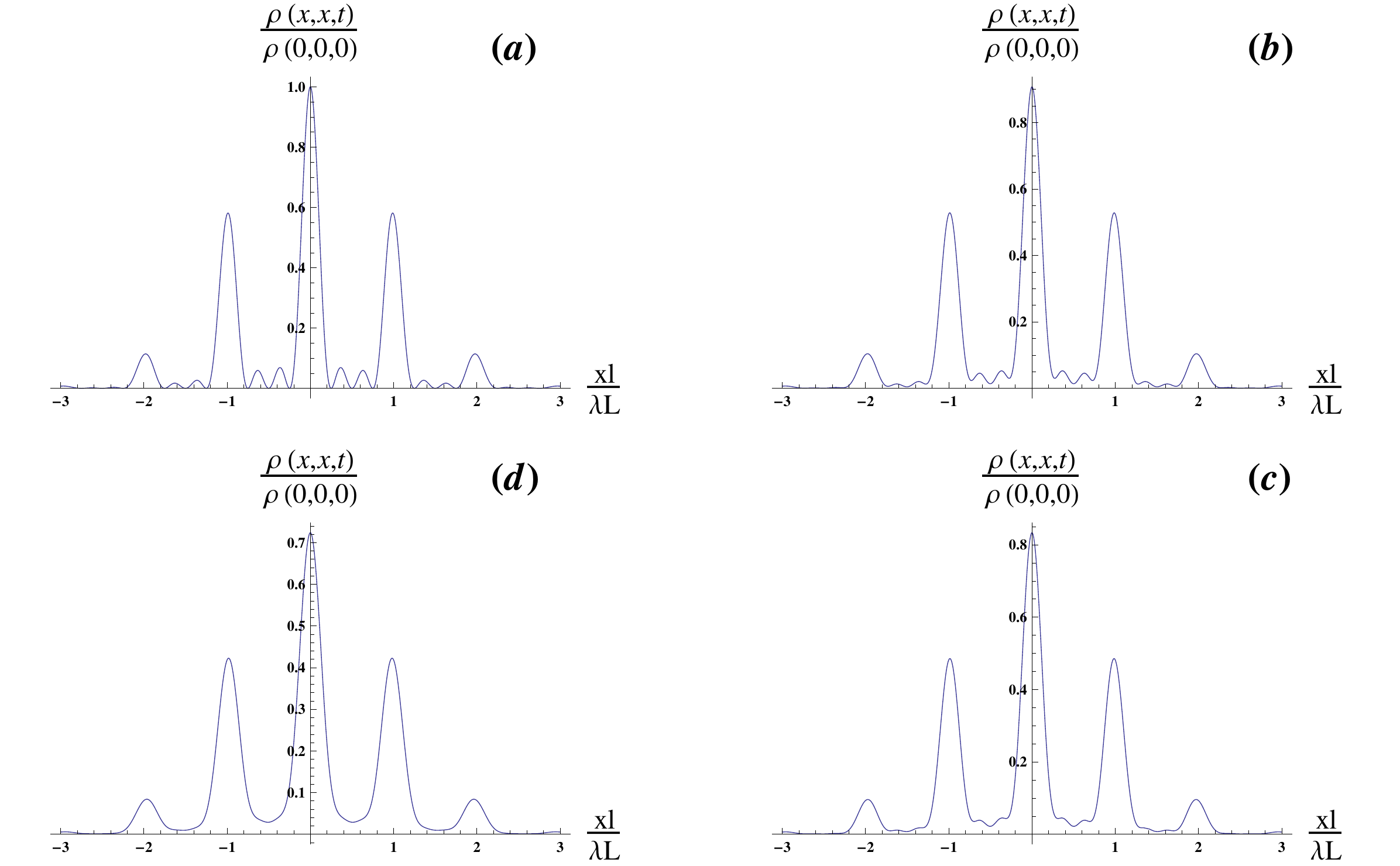}}}
\caption{Plots of $\rho(x,x,t)/\rho(0,0,0)$ at the screen, for 4-slit
interference, showing how decoherence progressively affects interference.
The following parameters used are adapted from the interference experiment
on ultracold Neon atoms \cite{neon}:
$m=3.349\times 10^{-26}~\text{Kg}, T = 2.5~\text{mK}, \lambda=0.018~\mu\text{m},
\ell=6~\mu\text{m}, L=37~\text{mm}$.
The plots are
for (a) $t/\tau_d=0$, (b) $t/\tau_d=1/24$, (c) $t/\tau_d=1/12$, (d) $t/\tau_d=1/6$,
where $\tau_{d} = 12\hbar^2/D \ell^2$.
As decoherence effects set in, the multi-slit nature of the interference pattern disappears,
and it is essentially reduced to a 2-slit interference pattern, as seen in (d).
}

\label{neon}
\end{figure}

Figures  \ref{neon} and \ref{c60} {illustrate} the effect  of decoherence {as described
by (\ref{rhoa}) for a simple example of four slits. { We plot the position probability  distribution  (\ref{rhoa}) (interference pattern) with  real parameters used in two significant experiments, the first by Shimizu  et al \cite{neon} reporting the observation of matter wave interference of ultracold Neon atoms and the second by Arndt et al \cite{c60} which observes matter wave interference of $C_{60}$ molecules. One can see that
decoherence leads to the washing}  out {of} the multi-slit  {character}
of the interference {pattern} first. Subsequently, the interference
pattern essentially reduces to that of a 2-slit interference pattern. This
aspect can be understood by recognizing that the strongest contribution to the decoherence comes from the two slits which are farthest apart, i.e., in this case slit number 1 and slit number 4. For interference between the amplitudes
from these two widely separated slits, quantum coherence has to be maintained over a
larger spatial distance. Since the time-scale over which
coherence is destroyed  between two spatially separated points is
inversely proportional to the square of their separation, the coherence between slit 1 and slit 4 in our example is destroyed faster. Since the weakest contribution to decoherence comes from the interfering amplitudes between nearest neighbour pairs of
slits, the four-slit pattern decays to a two-slit interference pattern
whose coherence is destroyed much more slowly. 
With the passage of time, decoherence eventually leads to a further loss in contrast and the eventual washing away of the fringes \cite{tq_av}(see Figure \ref{c60}). Figure 4 illustrates this for five-slit interference. In fact it can be shown that irrespective of the number of slits, all n-slit patterns, when affected by decoherence, degrade over time to a two-slit interference pattern which eventually washes away. It should be mentioned here that in this discussion, we are using the term coherence only in a qualitative sense, and not the one defined by
(\ref{coherence}).

\begin{figure}
\centerline{\resizebox{12cm}{!}{\includegraphics{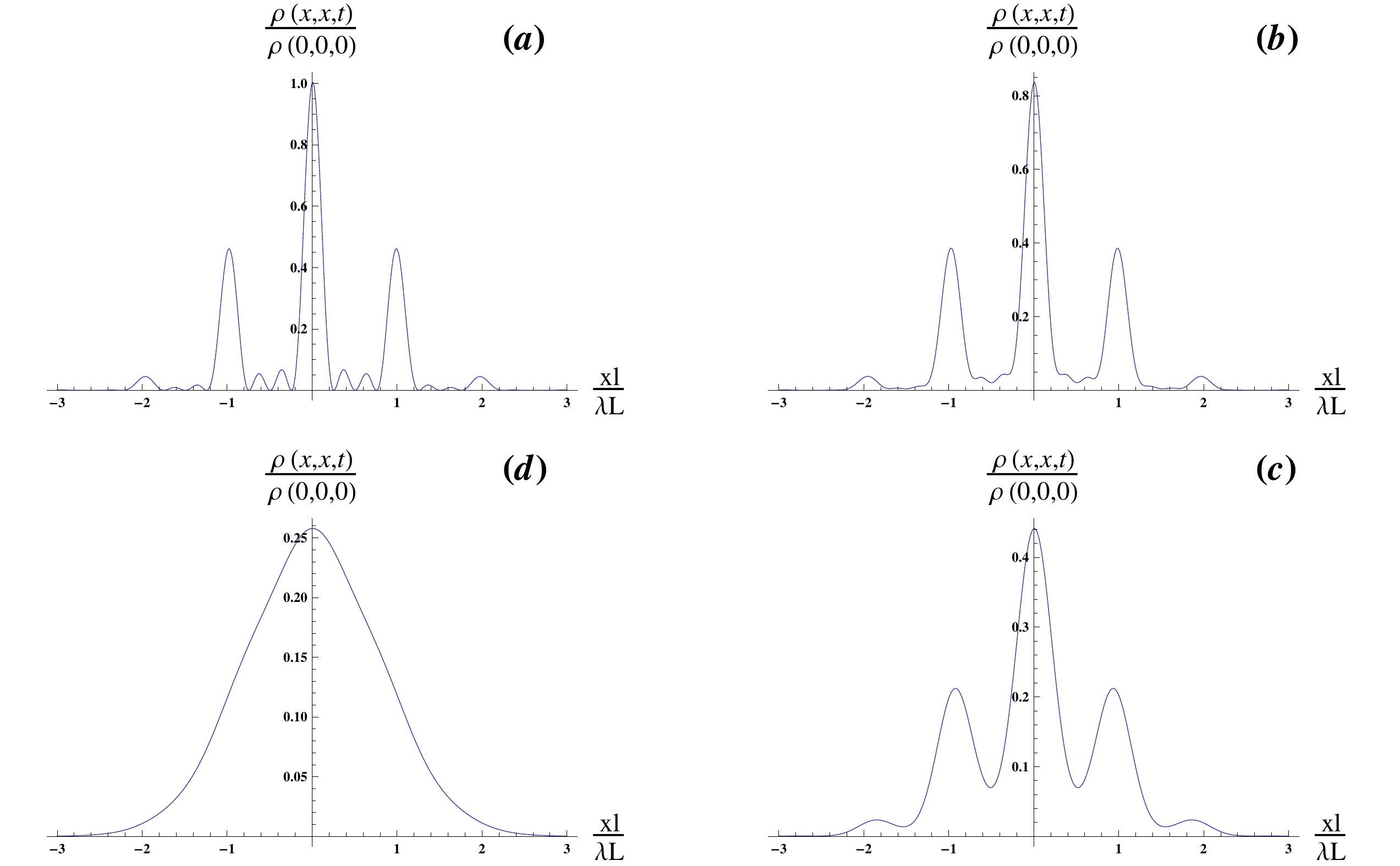}}}
\caption{Plots of $\rho(x,x,t)/\rho(0,0,0)$,
at the screen, for 4-slit
interference, showing how decoherence progressively affects interference.
The following parameters used are adapted from the interference experiment with
$C_{60}$ molecules \cite{c60}:
$m=1.2\times 10^{-24}~\text{Kg}, T = 900~\text{K}, \lambda=0.0025~\text{nm},
	\ell=100~\text{nm}, L=1.25~\text{m}$.
The plots are
for (a) $t/\tau_d=0$, (b) $t/\tau_d=1/12$, (c) $t/\tau_d=3/4$, (d) $t/\tau_d=4$,
where $\tau_{d} = 12\hbar^2/D \ell^2$.
As decoherence effects set in, the multi-slit nature of the interference  pattern disappears
first, and it is essentially reduced to a 2-slit interference pattern, as
seen in (b). With stronger decoherence effects, the effective 2-slit
interference pattern is also washed out.
}
\label{c60}
\end{figure}

\long\def\comment#1{}
\comment{
\begin{table*}
\caption{\label{tab:table1}Details and parameter values of experiments}
\begin{tabular}{cccc}
S.No & Parameter  & F. Shimizu \textit{et al.} \cite{neon} & M. Arndt \textit{et al.} \cite{c60}\\ \hline
1 & Source & ultracold Ne atoms  & Fullerene($C_{60}$) beam \\
2 & Mass ($m$) & $ 3.349 \times 10^{-26} $ Kg & $1.2 \times 10^{-24}$ Kg\\
3 & Temperature ($T$) & 2.5 $mK$ & 900 $K$ \\
4 & De broglie wavelength ($\lambda$) & $0.018$ $\mu m$  & $0.0025$ $nm$ \\ 
5 & Width of Gaussian ($ \epsilon $) & 1 $\mu m$ & 20 $nm$ \\
6 & Slit width ($d$) & 2 $\mu m$ & 50 $nm$ \\
7 & Distance between slits ($l$) & 6 $\mu m$ & 100 $nm$ \\
8 & Distance between slit and screen ($L$) & 37 $mm$ & 1.25 $m$ \\
9 & Fringe Width ($\beta$) & 111 $\mu m$ & 31250 $nm$ \\ 
\end{tabular}
\end{table*}
}

\subsection{Measuring coherence}

Next we turn to quantitatively extracting the coherence $\mathcal{C}$ \cite{nduality} as  defined in (\ref{coherence}) from the interference in (\ref{rho}) or (\ref{rhoa}).
In order to measure coherence from a n-slit interference, we need to have
such a path detector in place whose path-distinguishability is tunable
\cite{tania}.  The minimum requirement is that it should be switchable 
between two modes corresponding to (a) making all the paths completely
{\em indistinguishable} and (b) making all the paths fully
{\em distinguishable}. We denote  the two cases (a) and (b) by 
$\parallel$ and $\perp$, respectively. The procedure is as follows. First,
the intensity at a primary maximum $I_{max}^{\parallel}$ is measured when
the n paths are indistinguishable, i.e., $|d_i\rangle$s are all identical
and parallel. 
Next, the path-detector is switched to the mode (b) where all the n paths
are fully distinguishable, and the intensity $I_{max}^{\perp}$ is measured at
the same location on the screen as before. 

\begin{figure}
\centerline{\resizebox{12cm}{!}{\includegraphics{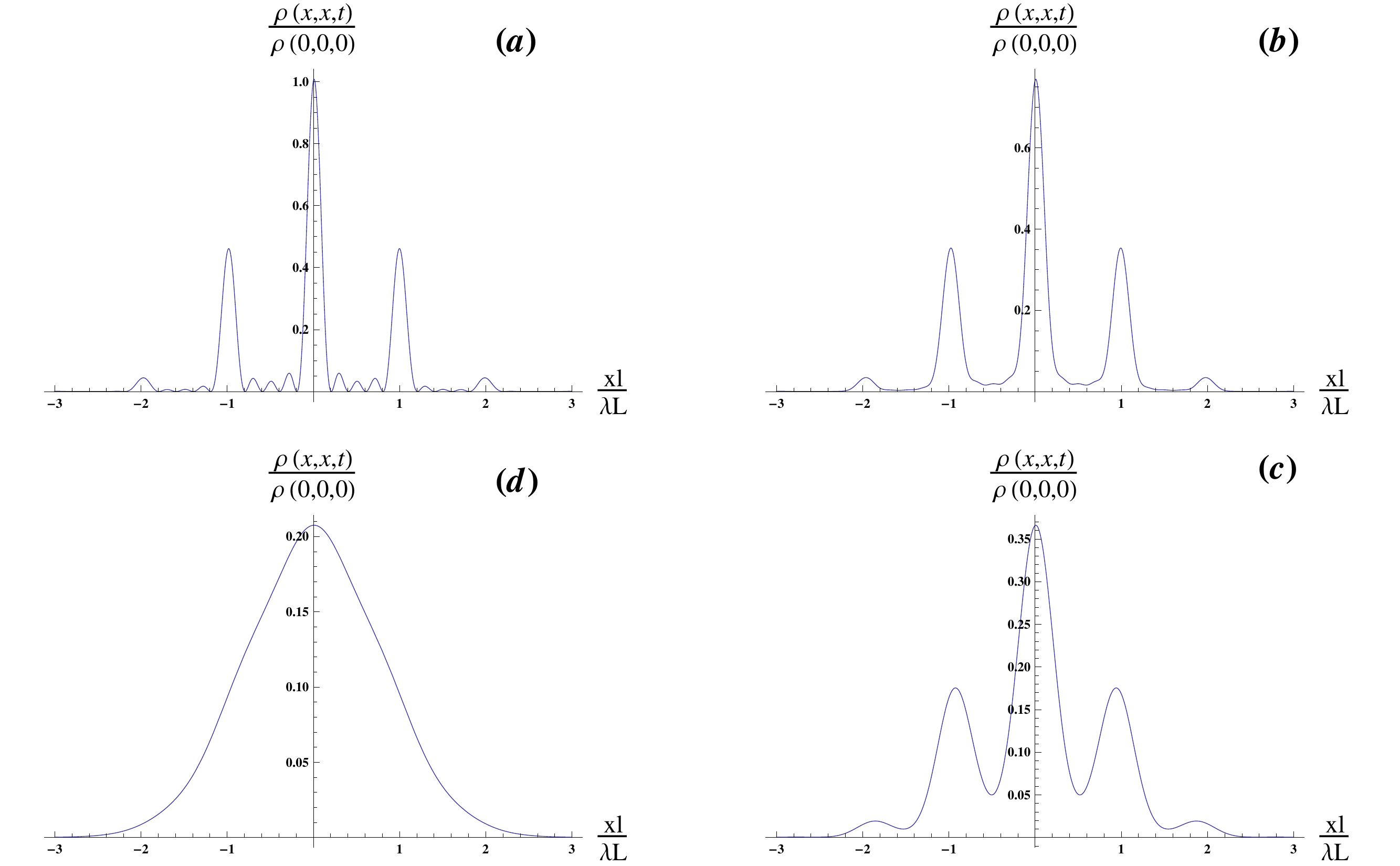}}}
\caption{Plots of $\rho(x,x,t)/\rho(0,0,0)$,
at the screen, for 5-slit
interference, showing how decoherence progressively affects interference.
The following parameters used are adapted from the interference experiment with
$C_{60}$ molecules \cite{c60}:
$m=1.2\times 10^{-24}~\text{Kg}, T = 900~\text{K}, \lambda=0.0025~\text{nm},
	\ell=100~\text{nm}, L=1.25~\text{m}$.
The plots are
for (a) $t/\tau_d=0$, (b) $t/\tau_d=1/12$, (c) $t/\tau_d=3/4$, (d) $t/\tau_d=4$,
where $\tau_{d} = 12\hbar^2/D \ell^2$.
As decoherence effects set in, the multi-slit nature of the interference pattern disappears
first, and it is essentially reduced to a 2-slit interference pattern, as
seen in (b). With stronger decoherence effects, the effective 2-slit
interference pattern is also washed out.
}
\label{c60n}
\end{figure}

Coherence of the incoming particles can then be measured as \cite{tania}
\begin{eqnarray}
\mathcal{C}_{expt} = \frac{1}{n-1}\frac{I_{max}^{\parallel} - I_{max}^{\perp}}{I_{max}^{\perp}} .
\label{Cexpt}
\end{eqnarray}
In order to extract $\mathcal{C}$ from (\ref{rhoa}), we assume that the
width of the slits $\epsilon$ is very narrow, and the narrow width Gaussians
$e^{{-(x-j\ell)^2}/{\epsilon^2}}$ in (\ref{rho0}), become very wide
$e^{-{2\epsilon^2(x-jl)^2/(\lambda L/\pi)^2}}$ in (\ref{rhoa}),
as the particle reaches the screen. This is the Fraunhofer limit
$\epsilon^2/\lambda L \ll 1$, and is satisfied by  the real experimental parameters 
used  in Figures \ref{neon}, \ref{c60} and \ref{c60n}.
Terms like
$e^{-{2\epsilon^2(x-jl)^2/(\lambda L/\pi)^2}}$ are wide Gaussians, whose
centers are shifted by tiny amounts of the order of a few slit-separations.
For all practical purposes, the values of all the Gaussians, at any point
$x$ on the screen, is almost the same, and thus independent of $j,k$.
We denote it by $g(x) \equiv e^{-{2\epsilon^2(x-jl)^2/(\lambda L/\pi)}}$.
The expression for $\rho(x,x,t)$ is then approximated as
\begin{eqnarray}
\rho(x,x,t) &\approx& \frac{g(x)}{\sqrt{\pi \alpha/2}}\left[\sum_{j=1}^n |c_j|^2
 + \sum_{j\ne k}
|c_jc_k||\langle d_k|d_j\rangle|{ e^{-(j-k)^2t/\tau_d}}\right. \nonumber\\
&&\left.\cos\left\{\frac{2\pi\ell(k-j)
(x-\ell\frac{k+j}{2})}{\lambda L } + \theta_k-\theta_j\right\}\right].
\label{rhoc}
\end{eqnarray}
Now,  in a n-slit interference, the primary maxima are at those points on the
screen where the value of all the cosine terms is 1, irrespective of the values of 
j, k. From (\ref{rhoc}) one can get the expressions for 
$I_{max}^{\parallel}$ and $I_{max}^{\perp}$ as follows:
\begin{eqnarray}
I_{max}^{\parallel} &=& \frac{g(x)}{\sqrt{\pi \alpha/2}}\left[\sum_{j=1}^n |c_j|^2
 + \sum_{j\ne k} |c_jc_k| e^{-(j-k)^2t/\tau_d} \right]\nonumber\\
I_{max}^{\perp} &=& \frac{g(x)}{\sqrt{\pi \alpha/2}}\left[\sum_{j=1}^n |c_j|^2
\right].
\label{Imax}
\end{eqnarray}
Coherence can then be calculated using (\ref{Cexpt}):
\begin{eqnarray}
\mathcal{C} &=& \frac{1}{n-1} \sum_{j\ne k} |c_jc_k| {e^{-(j-k)^2t/\tau_d}}\nonumber\\
&=& \frac{1}{n-1} \sum_{j\ne k} |c_jc_k| \exp\left\{\frac{-(j-k)^2\ell^2 tm\gamma k_BT}{6\hbar^2}\right\},
\label{Ccal}
\end{eqnarray}
where we have used the property $\sum_{j=1}^n |c_j|^2=1$.
Eqn. (\ref{Ccal}) is interesting  as it quantitatively captures the coherence of {the quanton}  in terms of  {real} parameters of the environment and its coupling.
For example, it quantifies the effect of increasing or decreasing the
temperature of the heat-bath on the coherence of the particle. This
feature can be experimentally tested in matter wave interference
experiments by  changing the temperature of the ambient gas reservoir
\cite{horn}.
Note that while the interference pattern captured by (\ref{rhoa}) depends
on the overlaps of the path-detector states, the coherence represented by
(\ref{Ccal}),
is completely independent of the path detector.

The two-slit ($n=2$) case is particularly useful, as we shall see in the
following. For $n=2$, (\ref{Ccal}) reduces to
\begin{eqnarray}
\mathcal{C} &=&  2|c_1c_2| {e^{-t/\tau_d}}.
\label{C2}
\end{eqnarray}
This immediately allows us to represent the decoherence time, $\tau_d$, in terms
of the coherence , $\mathcal{C}$, as
\begin{eqnarray}
{\tau_d} = \frac{t}{\log(2|c_1c_2|/\mathcal{C})},
\label{td0}
\end{eqnarray}
or
\begin{eqnarray}
{\tau_d} = \frac{\lambda Lm/h}{\log(2|c_1c_2|/\mathcal{C})}.   
\label{td}
\end{eqnarray}
The above expression can be extremely useful for the following reason. 
In a two-slit interference experiment, if one can experimentally measure
coherence using (\ref{Cexpt}), it allows one to  determine the
decoherence time.   {For instance, one may be interested in knowing the 
decoherence time-scale in a particular experimental situation, limited by
the degree of vacuum and other constraints.} For example, if one
can hook up a {\em symmetric} two-slit interference experiment, with which-way
detection, it allows one to experimentally determine the decoherence
time-scale simply by using the following relation:
\begin{eqnarray}
{\tau_d} = \frac{\lambda Lm/h}{\log\left(\frac{I_{max}^{\perp}}{I_{max}^{\parallel} - I_{max}^{\perp}}\right)}.
\label{td1}
\end{eqnarray}
Thus decoherence can be monitored by measuring intensities in the interference
pattern. While considerable progress in the theoretical understanding
of decoherence has been made in recent times,  experiments that can
control environmental coupling and monitor progressive decoherence
are few \cite{monroe, exptdecoh1}. In particular, measurement of
the decoherence time in various situations remains a big challenge.
Knowledge of the decoherence time is crucial in systems where quantum
coherence is exploited for  information processing and computing, making
(\ref{td1}) an extremely useful result.

\section{Conclusions}

We have theoretically analyzed a multi-slit interference with which-way
detection, in the presence of environment-induced decoherence.
Decoherence degrades the interference in an interesting manner, with the
multi-slit features disappearing first, and the interference pattern reducing to
an effective decohering two-slit interference pattern, which eventually washes out. An analytical expression for the coherence of the interfering particle is obtained in terms of the parameters of the environment and the particle-environment coupling. Using that we show that decoherence can be quantified by measuring coherence. For the particular case of two-slit interference, we show that the decoherence time-scale can be experimentally determined by measuring coherence, which in turn, can be determined by measuring interference intensities
in a particular way.This procedure may become useful in developing methods of monitoring decoherence in many potentially useful quantum systems which exploit quantum coherence.

\section*{Acknowledgement}
Sandeep Mishra thanks Guru Gobind Singh Indraprastha Univeristy for the Indraprastha Research Fellowship.

\end{document}